\documentclass[iop]{emulateapj}
\usepackage{color}

\def\lup1{Lupus\,I}

\shorttitle{Magnetic Fields in \lup1}
\shortauthors{G.~A.~P. Franco \& F.~O. Alves}

\begin{document}

\title{Tracing the magnetic field morphology of the Lupus\,I molecular cloud{\large $^\star$}}

\author{G.~A.~P. Franco} 
\affil{Departamento de F\'\i sica -- ICEx -- UFMG, Caixa Postal 702, 30.123-970
Belo Horizonte, Brazil; franco@fisica.ufmg.br}

\and 

\author{F.~O. Alves}
\affil{Max-Planck-Institut f\"ur extraterrestrische Physik, Giessenbachstr. 1, D-85748 
Garching, Germany; falves@mpe.mpg.de}

\altaffiltext{$^\star$}{Based on observations collected at Observat\'orio do Pico dos 
Dias, operated by Laborat\'orio Nacional de Astrof\'\i sica (LNA/MCTI, Brazil).}

\begin{abstract}
Deep $R$-band CCD linear polarimetry collected for fields with lines-of-sight 
toward the \lup1 molecular cloud is used to investigate the properties of the magnetic 
field within this molecular cloud. The observed sample contains about 7000 stars, almost 
2000 of them with polarization signal-to-noise ratio larger than 5. These data cover 
almost the entire main molecular cloud and also sample two diffuse infrared patches 
in the neighborhood of \lup1. The large scale pattern of the plane-of-sky projection of the 
magnetic field is perpendicular to the main axis of \lup1, but parallel to the two diffuse infrared 
patches. A detailed analysis of our polarization data combined with the 
{\it Herschel}/SPIRE 350\,$\mu$m dust emission map shows that the principal 
filament of \lup1 is constituted by three main clumps acted by magnetic fields having different 
large-scale structure properties. These differences may be the reason for the observed
distribution of pre- and protostellar objects along the molecular cloud and its apparent
evolutive stage. On the other hand, assuming that the magnetic field is composed by a 
large-scale and a turbulent components, we find that the latter is rather similar in all 
three clumps. The estimated plane-of-sky component of the large-scale magnetic field 
ranges from about 70\,$\mu$G to 200\,$\mu$G in these clumps. The intensity increases 
towards the Galactic plane. The mass-to-magnetic flux ratio is much smaller than unity, 
implying that \lup1 is magnetically supported on large scales.
\end{abstract}

\keywords{ISM: clouds -- ISM: individual objects: \lup1 -- 
ISM: magnetic fields -- Techniques: polarimetry}

\section{Introduction}\label{int}

The shape of an interstellar cloud may provide a wealth of information on the mechanisms that 
govern cloud formation and evolution. In the past decades, there were accumulated evidences 
that the interstellar medium (ISM) has a filamentary appearance. The {\it IRAS} all sky survey 
revealed a profusion of parsec-scale filaments in the diffuse ISM \citep{low}, while observations 
of dust emission, interstellar reddening, and molecular gas emission have shown 
that filaments are also omnipresent in the Galactic molecular clouds 
\citep[see][for a recent review on the molecular ISM]{HF12}. This outstanding
characteristic of the ISM was highlighted by the recent results from the
{\it Herschel} satellite's far-infrared survey of the inner Galactic plane 
\citep[e.g.,][]{MSB10}. On the other hand, a comprehensive analysis of the relative 
orientation between filamentary interstellar structures and the Galactic magnetic field 
performed across the whole sky using {\it Planck} data, has shown that the diffuse 
interstellar matter is preferentially aligned with the magnetic field, while molecular clouds 
appear perpendicularly aligned \citep{AAA14}. Thus, magnetic fields must 
play an important role in the ISM molecular clouds formation and evolution; 
nevertheless, understanding its extent and importance in star formation and the 
competition between magnetic, and turbulent forces is one of the most heated 
debate in modern astrophysics \citep[e.g.,][]{MC99, MK04, MO07, Cr12}. 
Although some theoretical investigations point to supersonic turbulence as the 
prevailing, possibly the most important, mechanism in the formation of structure 
and evolution of molecular clouds, most recent results like the one obtained by
the {\it Planck} group and others \citep[e.g.,][]{LDG09, LFH13, VSB11, 
SHM12, VKZ14} testify that the magnetic field is also of crucial importance. 

Most of what we know about the plane-of-sky component of the Galactic 
magnetic field comes from the observation of starlight which is polarized by interstellar 
dust grains in the foreground. These grains are elongated and according to the pioneering 
model proposed by \citet{DG51} they would be aligned by the magnetic field perpendicular 
to the field lines. Nevertheless, the paramagnetic relaxation mechanism 
proposed by \citeauthor{DG51} exhibits difficulty to reconcile with observations 
and the actual mechanism responsible for this process has proven to be one of the longest 
standing problems in astrophysics \citep[see, for instance,][for a comprehensive review on 
the dust alignment theory]{L07}. Recent advances in this area suggest that radiative torque
alignment, originally proposed by \citet{DM76}, has become the favored mechanism to 
explain grain alignment, as showed by models \citep[e.g.,][]{HL14, HLA15} and
observations \citep[e.g.,][]{AFG14, JBK15}.The same aligned dust grains that are 
responsible for dichroic absorption in the visible and near-infrared wavelengths can 
emit polarized thermal radiation in the mid- and far-infrared, and radio wavelengths.  

The main purpose of this investigation is to study the morphology of the magnetic field in
the vicinity of the elongated \lup1 molecular cloud, the most massive cloud 
\citep[$\sim$1200\,M$\sun$;][]{TDM96} in the Lupus complex, a loose conglomerate of dark
clouds in the solar vicinity \citep[distant $\sim$150\,pc according to][]{CIA00,FR02, AF06,
LLA08}. Many young objects from pre-stellar to T\,Tauri class stars have been associated to 
these clouds, making them one of the nearest known low-mass star forming regions 
\citep[for a comprehensive review on the Lupus dark clouds complex see][]{FC08}. 
\lup1 is particularly interesting because it is the youngest cloud in the complex \citep{BPB12}.
This hypothesis is reinforced by the recent census of pre-stellar sources conducted by 
\citet{RBS13} which suggests that \lup1 is currently undergoing a large star formation
event. 

Earlier references on polarimetric investigations concerning \lup1 can be found in the 
literature \citep[][]{MG91, RMA98, MAA14}. The former two studies are based on optical 
linear polarization data, while the latter concerns submm polarization data. All of them 
conclude that in large scale the plane-of-sky component of the magnetic field is 
perpendicularly aligned to the main axis of  \lup1, like what is observed 
for many others filamentary molecular clouds \citep[e.g.,][]{HVS87, PM04, AFG08}. In this 
paper we introduce the results from CCD image polarimetry conducted in the $R$-band for 
fields with lines-of-sight covering almost the whole area of \lup1. These data 
allowed us to go beyond previous works and obtain a detailed description of the configuration of 
the magnetic field permeating the \lup1 molecular cloud. 

\section{Observations}\label{obs}

The polarimetric data were collected with the 1.6\,m and the IAG 60\,cm 
telescopes at Observat\'orio do Pico dos Dias (LNA/MCTI, Brazil) in missions 
conducted in several nights from 2004 to 2008. The data were obtained with 
the use of a specially adapted CCD camera to allow polarimetric measurements 
--- for a suitable description of the polarimeter see \citet{AM96}. $R$-band 
polarimetry was obtained for 62 fields (with field of view of about $12\arcmin 
\times 12\arcmin$ each) covering the main body of the Lupus\,I molecular
cloud and two neighboring diffuse infrared patches. The lines-of-sight 
of these fields were chosen taking as reference the {\it IRAS} 100\,$\mu$m 
emission map of the region.  Among the 62 sky positions, 27 were 
observed at the IAG 60\,cm telescope, and the remaining 35 fields were observed at 
the 1.6\, m telescope. Some of the observed fields overlapped each 
other, either totally or partially, so that the covered effective area is smaller than 
the total area covered by the individual frames. In both telescopes the integration time 
was set to 120\,s for each frame, being that at the IAG 60\,cm telescope five frames were
collected and co-added for each position of the half-wave plate (totalizing
600\,s per waveplate position), while at the 1.6\,m telescope only one frame
was obtained for each waveplate position. Most of these observations shared telescope 
time with a project dedicated to the Pipe Nebula, whose results were previously 
published elsewhere \citep{FAG10}. The interested reader is referred to that paper 
for details on the data reduction, as well as on the observed unpolarized and 
polarized standard stars used to check for any possible instrumental polarization and
for determining the reference direction for the position angles, respectively.
The normalized linear polarization is calculated from a least-square
solution, which yields the degree of polarization ($P$), the polarization
position angle ($\theta$, measured from north to east), the Stokes parameters
$Q$ and $U$, as well as the theoretical (i.e., the photon noise) and measured errors,
$\sigma_{\rm theo}$ and $\sigma_{\rm fit}$, respectively. The latter are 
obtained from the residuals of the observations at each waveplate
position angle ($\psi_i$) with respect to the expected $\cos\, 4\psi_i$ curve \citep[see][for 
a brief description on the method used to obtain the least-square solution for the Stokes 
parameters and the associated errors]{MBR84}.

\begin{figure}
\epsscale{1.2}
\plotone{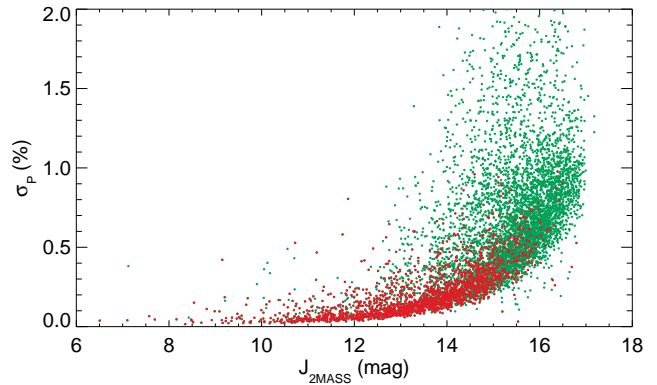}
\caption[]{Distribution of the polarimetric errors as a function of the 2MASS $J$-band magnitude.
The red dots represent stars with $P/\sigma_P \ge 5$, while the green dots represent the remaining 
stars. The distribution shows characteristics of estimated errors dominated by photon shot noise.}
\label{sigma_x_j}
\end{figure}

\begin{deluxetable*}{ccccccccrcccccc}
\tablecolumns{15}
\tablewidth{515pt} 
\tablecaption{Sample of the polarimetric data obtained for \lup1}
\tablenum{1}
\label{tab_data}
\tablewidth{0pt}
\tablehead{\colhead{ID} & \colhead{$\alpha_{2000}$} & \colhead{$\delta_{2000}$} & \colhead{$l$} & \colhead{$b$} & 
\colhead{$P$} &
\colhead{$\sigma_{\rm fit}$} & \colhead{$\sigma_{\rm theo}$} & \colhead{$\theta$} & \colhead{$J$} &
\colhead{$\sigma_J$}  & \colhead{$H$} & \colhead{$\sigma_H$} & 
\colhead{$K_S$} & \colhead{$\sigma_{K_S}$} \\ 
\colhead{} & \colhead{($^{\rm h}$ \phn $^{\rm m}$ \phn $^{\rm s}$)} & \colhead{($\degr \phn\phn ' \phn \phn ''$)} & \colhead{(\degr)} & \colhead{(\degr)} & 
\colhead{(\%)} & \colhead{(\%)} & \colhead{(\%)} & \colhead{(\degr)} & \colhead{(mag)} & \colhead{(mag)} &
\colhead{(mag)} & \colhead{(mag)} & \colhead{(mag)} & \colhead{(mag)} } 
\startdata
  \phn1 & 15\phn36\phn13.79 & $-$33\phn02\phn32.4 & 338.4735 & 18.2235 & 0.847 & 0.132 & 0.075 &  89.6 & 12.145 & 0.024 & 11.518 & 0.021 & 11.392 & 0.024 \\
  \phn2 & 15\phn36\phn15.85 & $-$33\phn04\phn10.7 & 338.4617 & 18.1976 & 1.114 & 0.411 & 0.342 & 123.9 & 14.775 & 0.039 & 14.172 & 0.044 & 14.169 & 0.054 \\
  \phn3 & 15\phn36\phn15.86 & $-$33\phn10\phn47.0 & 338.3902 & 18.1110 & 1.585 & 0.186 & 0.215 & 102.2 & 13.814 & 0.024 & 13.291 & 0.021 & 13.134 & 0.030 \\
  \phn4 & 15\phn36\phn15.98 & $-$33\phn08\phn31.0 & 338.4151 & 18.1404 & 1.263 & 0.010 & 0.071 &  85.7 & 11.939 & 0.024 & 11.294 & 0.023 & 11.085 & 0.021 \\
  \phn5 & 15\phn36\phn16.07 & $-$33\phn02\phn32.2 & 338.4801 & 18.2187 & 0.526 & 0.353 & 0.417 &  21.2 & 15.131 & 0.049 & 14.657 & 0.046 & 14.607 & 0.084 \\
  \phn6 & 15\phn36\phn16.22 & $-$33\phn07\phn46.7 & 338.4238 & 18.1496 & 1.396 & 0.441 & 0.365 &  99.8 & 14.915 & 0.033 & 14.321 & 0.044 & 14.150 & 0.061 \\
  \phn7 & 15\phn36\phn16.28 & $-$33\phn06\phn25.2 & 338.4387 & 18.1673 & 1.201 & 0.215 & 0.456 & 128.0 & 15.387 & 0.049 & 14.937 & 0.079 & 14.823 & 0.104 \\
  \phn8 & 15\phn36\phn16.29 & $-$33\phn06\phn45.5 & 338.4350 & 18.1628 & 1.546 & 0.436 & 0.445 &  91.5 & 15.105 & 0.040 & 14.560 & 0.063 & 14.443 & 0.075 \\
  \phn9 & 15\phn36\phn16.78 & $-$33\phn04\phn42.1 & 338.4587 & 18.1887 & 0.494 & 0.293 & 0.211 &  99.9 & 13.935 & 0.029 & 13.321 & 0.030 & 13.186 & 0.033 \\
 10 & 15\phn36\phn16.88 & $-$33\phn06\phn47.6 & 338.4364 & 18.1611 & 0.697 & 0.147 & 0.146 & 109.4 & 13.753 & 0.030 & 13.339 & 0.030 & 13.205 & 0.034  
  \enddata
\tablecomments{Table 1 is only available in its entirety at the CDS via anonymous ftp to 
cdsarc.u-strasbg.fr (130.79.128.5) or via http://cdsarc.u-strasbg.fr/viz-bin/qcat?J/ApJ/XXX/XXX. A 
portion is shown here for guidance regarding its content. Columns, respectively, represent the 
star's identifier number in our catalogue (ID), the equatorial coordinates ($\alpha$, $\delta$) retrieved from 
the 2MASS catalogue, the Galactic coordinates ($l$, $b$), the polarization degree (P), together with 
the two estimated errors for this quantity (the measured ($\sigma_{\rm fit}$) and the theoretical 
($\sigma_{theo}$) errors), the polarization angle ($\theta$), and the 2MASS magnitudes ($J$, $H$, $K_S$) 
with their associated uncertainties ($\sigma_J$, $\sigma_H$, $\sigma_{K_S}$), respectively.}
\end{deluxetable*}

The astrometric calibration was done in two steps. A first solution for each observed 
frame was obtained from the few stars identified from the Second Digitized Sky Survey
(DSS2 red). Such preliminary solution was used to identify stars from the 2MASS 
catalogue with lines-of-sight toward the frames, and these stars were used to provide 
a final astrometric solution. This procedure provided accurate coordinates and 
automatically cross correlated our objects with the 2MASS data catalogue. 

\begin{figure}
\epsscale{1.15}
\plotone{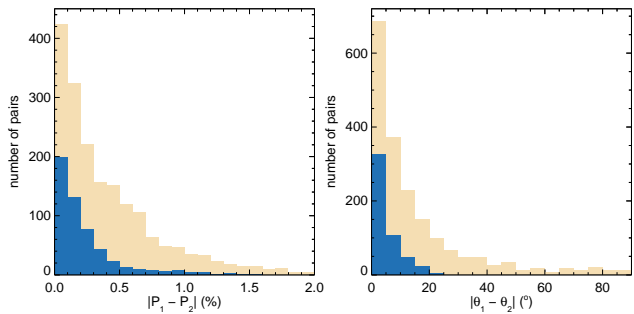}
\caption[]{Distribution of the absolute differences in degree of polarization (left) and polarization 
angles (right) of pairs of measurements for the same star. The wheat histograms give the distribution 
for all pairs, while the blue histograms are for pairs presenting $P/\sigma_P \ge 5$ only.}
\label{common}
\end{figure}

Our final polarimetric sample contains about 7\,000 stars, 1938 of which have 
$P/\sigma_P \ge 5$, where $\sigma_P = \max \{ \sigma_{\rm fit},\sigma_{\rm theo} \}$.
In Table\,\ref{tab_data} we introduce the first 10 lines of our data catalogue 
for guidance regarding its form and content\footnote{Table\,\ref{tab_data}
is only available at the CDS via anonymous ftp to cdsarc.u-strasbg.fr (130.79.128.5) or via
http://cdsarc.u-strasbg.fr/viz-bin/qcat?J/ApJ/XXX/XXX}.
Figure\,\ref{sigma_x_j} gives the distribution of the estimated polarimetric errors,
$\sigma_P$, as a function of the $J_{\rm 2MASS}$ magnitude. A comparison between 
data obtained for stars in common to more than one observed frame has shown 
that the agreement for the measured polarization degree as well as for
the position angle were very good. Fig.\,\ref{common} presents the obtained distribution
of the absolute difference between two measurements for the degree of polarization (left panel), 
$|P_1- P_2|$, and polarization angles (right panel), $|\theta_1 - \theta_2|$. The histograms 
coloured in wheat represent the distribution for all observed pairs while the histograms coloured in
blue are for pairs of measurements having 
$P/\sigma_P \ge 5$. Taking into account the latter group we see that the majority of the 
obtained pairs presents a difference in polarization angle that is $\le 5\deg$, which is 
consistent with the expected uncertainty for measurements of this quantity 
\citep[$\sigma_\theta = 28\fdg65\,\sigma_P/P$,][]{SK74}. It is also seen, for the latter
group, that for the majority of the observed pairs the measured difference in degree of 
polarization is $\le 0.2\%$.

\begin{figure}
\epsscale{1.2}
\plotone{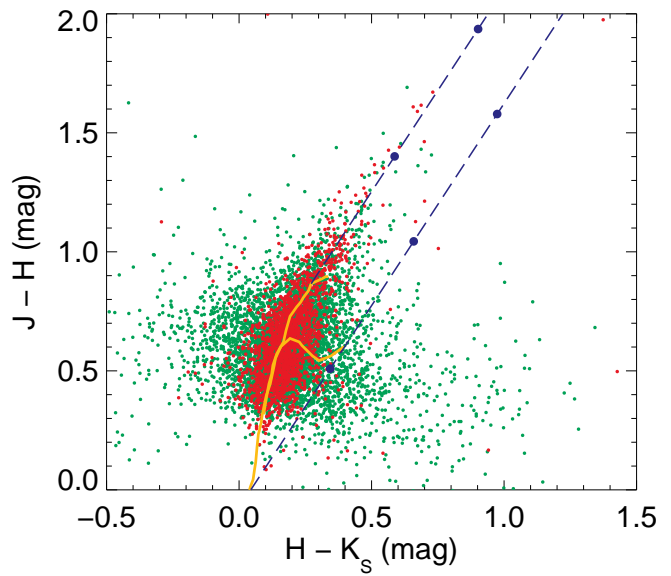}
\caption[]{2MASS ($J-H$) $\times$ ($H-K_{\rm S}$) color-color distribution of the observed stars. The 
solid curves represent the locus of main-sequence and giant stars and the diagonal dashed lines delimit 
the normal reddening zone; the large dots show intervals of $A_V = 5$ mag. We used the same color code 
as in Fig.\,\ref{sigma_x_j} for representing the data points.}
\label{jh_vs_hk}
\end{figure}

\begin{figure*}
\plotone{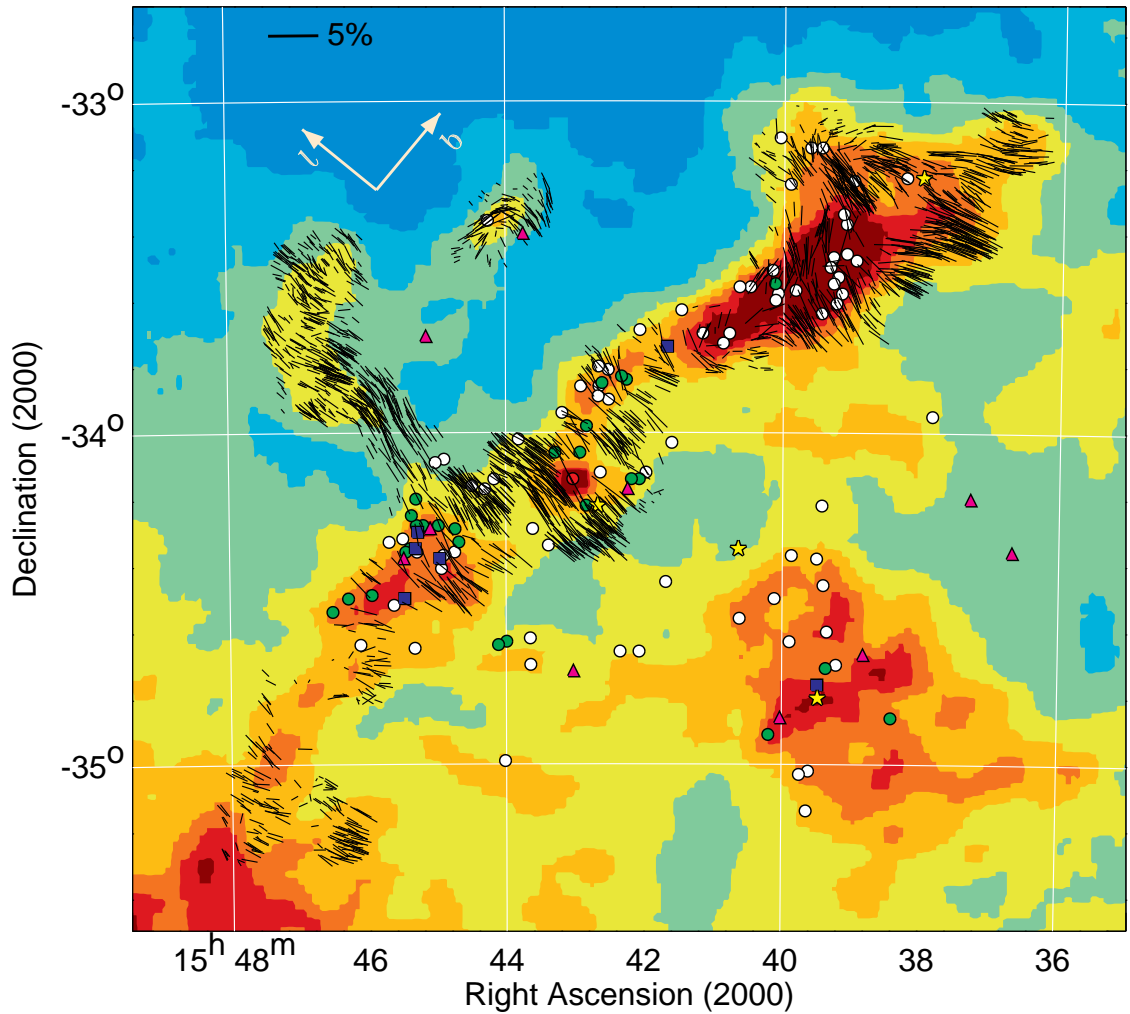}
\caption[]{The obtained polarization degree vectors overlaid on the {\it IRAS} 100\,$\mu$m 
emission map of Lupus\,I and its surrounding region. The length of each vector correlates linearly 
with the measured degree of polarization according to the scale indicated in the upper left corner, 
and its orientation gives the direction of the plane-of-sky component of the local magnetic field. The 
position of different classes of young stellar objects, from \citet{RBS13}, are indicated according with 
the following code: unbound cores (white circles), prestellar cores (green circles), class 0 (red circle), 
class I (triangles), class II (squares), and class III (stars). Directions of increasing Galactic longitude 
({\it l}\,) and latitude ({\it b}) are shown in the upper left corner.
}
\label{iras_map}
\end{figure*}

It is instructive to analyze the 2MASS color-color diagram obtained for the observed
stars, shown in Fig.\,\ref{jh_vs_hk}. In this diagram the  red dots represent stars with 
$P/\sigma_P \ge 5$, while stars with lower polarimetric measurement quality are 
represented by the green dots. A large number of stars belonging to this latter group present color 
indices which are somehow peculiar; it must be noticed, however, that most of them are
faint and also have lower photometric quality in the 2MASS catalogue. On the other hand,
most of the stars with $P/\sigma_P \ge 5$ have the highest photometric quality in that
catalogue. An information we can visually obtain from the 2MASS color-color diagram is that 
most of the stars observed by our polarimetric survey having $P/\sigma_P \ge 5$ are, in general, 
probing the outskirts of the dense molecular cloud, with $A_V \le 5$ mag. \citet{Cam99b} 
has estimated peak extinction towards \lup1 of about $A_V \simeq 7.1$ mag, nevetheless more
recent estimates of extinction in this cloud indicates $A_V > 20$ mag towards the darkest
areas \citep{BPB12}. 

\section{Data Analysis}\label{ana}
\subsection{Large scale morphology of the magnetic field}\label{large_scale}

Hereinafter, only stars having $P/\sigma_P \ge 5$ will be used in our analysis. 
Figure\,\ref{iras_map} introduces the distribution of the observed polarization 
vectors overlaid on the {\it IRAS} 100\,$\mu$m emission map of \lup1 and its 
surroundings. In order to complement the information on the surveyed area, we 
also provide the positions of the different classes of young stellar objects retrieved 
from \citet{RBS13}. 

The long axis of the main filament of \lup1 is roughly aligned at an angle of 127\degr\ 
east of north, resulting to a normal at about 37\degr, and at first glance 
Fig.\,\ref{iras_map} shows that the large scale plane-of-sky projection 
of the magnetic field appears to be perpendicularly oriented to this main axis, 
reinforcing the already mentioned previous results \citep[][]{MG91, RMA98, MAA14}. 
Nevertheless, a more detailed inspection suggests the existence of patterns, 
mainly at the north-western part of the cloud, which is also the region presenting 
the strongest {\it IRAS} 100\,$\mu$m emission. Close to the central part of the 
cloud, that is, around $15^{\mathrm h}43^{\mathrm m}$, $-34\arcdeg10\arcmin$, 
the polarization vectors became more uniformly aligned, while in the south-eastern 
surveyed part of the complex the degree of polarization seems to decrease, compared 
to the previous regions.

\begin{figure}
\epsscale{1.1}
\plotone{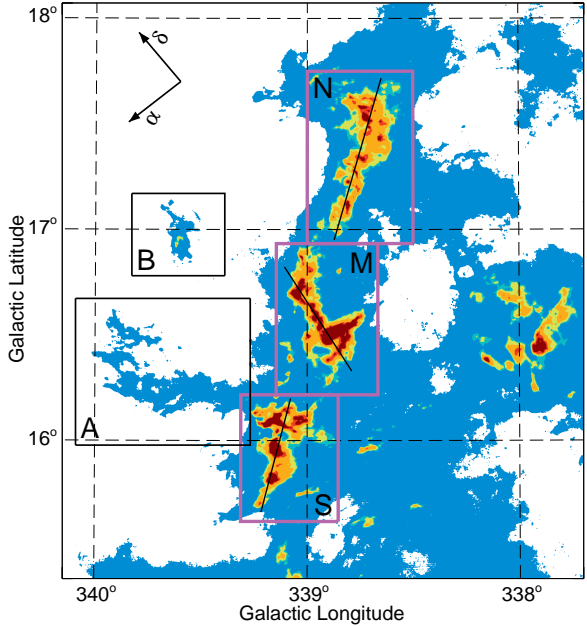}
\caption[]{The {\it Herschel}/SPIRE 350\,$\mu$m dust emission map for \lup1. The main
filament is dominated by three regions of enhanced emission delimited by the rectangles 
(colored purple), designated as the ``North'', ``Middle'', and ``South'' clumps according to 
their Galactic latitude. The black lines roughly indicate the direction of the filaments. The two 
secondary infrared emission patches are delimited by the boxes identified by ``A'' and ``B''. 
Directions of increasing right ascension ($\alpha$) and declination ($\delta$) are shown in 
the upper left corner.}
\label{herschel_gal}
\end{figure}

The obtained distribution of the polarization angles has characteristics of a 
normal distribution centered at 49\degr, that is, about 12\degr\ to the east of the normal 
to the main axis of  \lup1. The distribution is relatively broad with a standard deviation 
$\sigma_\theta \simeq 19\fdg6$, corresponding to a FWHM $\approx 46\degr$. The 
obtained values for this distribution agrees perfectly well with the one previously reported 
by \citet{MG91}, which were obtained from a much smaller sample (98 stars) and
\citet{RMA98}, who obtained a mean polarization angle of 52\degr\ with a
standard deviation $\sigma_\theta = 23\degr$, for stars close to the rim of \lup1.

\subsection{Polarization properties}\label{pol_theta}

When viewed in Galactic coordinates, the main body of the \lup1 molecular cloud is almost 
aligned in the north-south direction. We take advantage of this fact to perform a detailed 
analysis of the distribution of polarization degree and position angles over the main 
filament and the surrounding infrared patch clouds. \lup1 was observed 
as part of the {\it Herschel} Gould Belt survey \citep{AS05}, and although the mapped area is 
slightly smaller than the one covered by our polarimetric sample, it has a higher angular resolution 
than the {\it IRAS} maps and will be useful for the following analysis. In 
Fig.\,\ref{herschel_gal} we display the {\it Herschel}/SPIRE 350\,$\mu$m dust emission map 
obtained for \lup1\footnote{Retrieved from the {\it Herschel Gould Belt Survey Archives} at 
http://gouldbelt-herschel.cea.fr/} reprojected in Galactic coordinates. Five regions of interest 
are delimited by the boxes in this Figure. Three rectangles in the main \lup1 filament delimit 
regions of enhanced infrared emission that will be referred to as the ``North'', ``Middle'', and 
``South'' clumps, respectively from north to south. Each of these clumps are elongated, being 
that the two externals present almost the same direction as the whole filament 
($\theta_{\rm gal} \approx 164\degr$, east of north), while the middle one, at 
$\theta_{\rm gal} \approx 33\degr$, is tilted by almost 50\degr\ in the direction from the 
north to east.

\begin{figure}
\epsscale{1.1}
\plotone{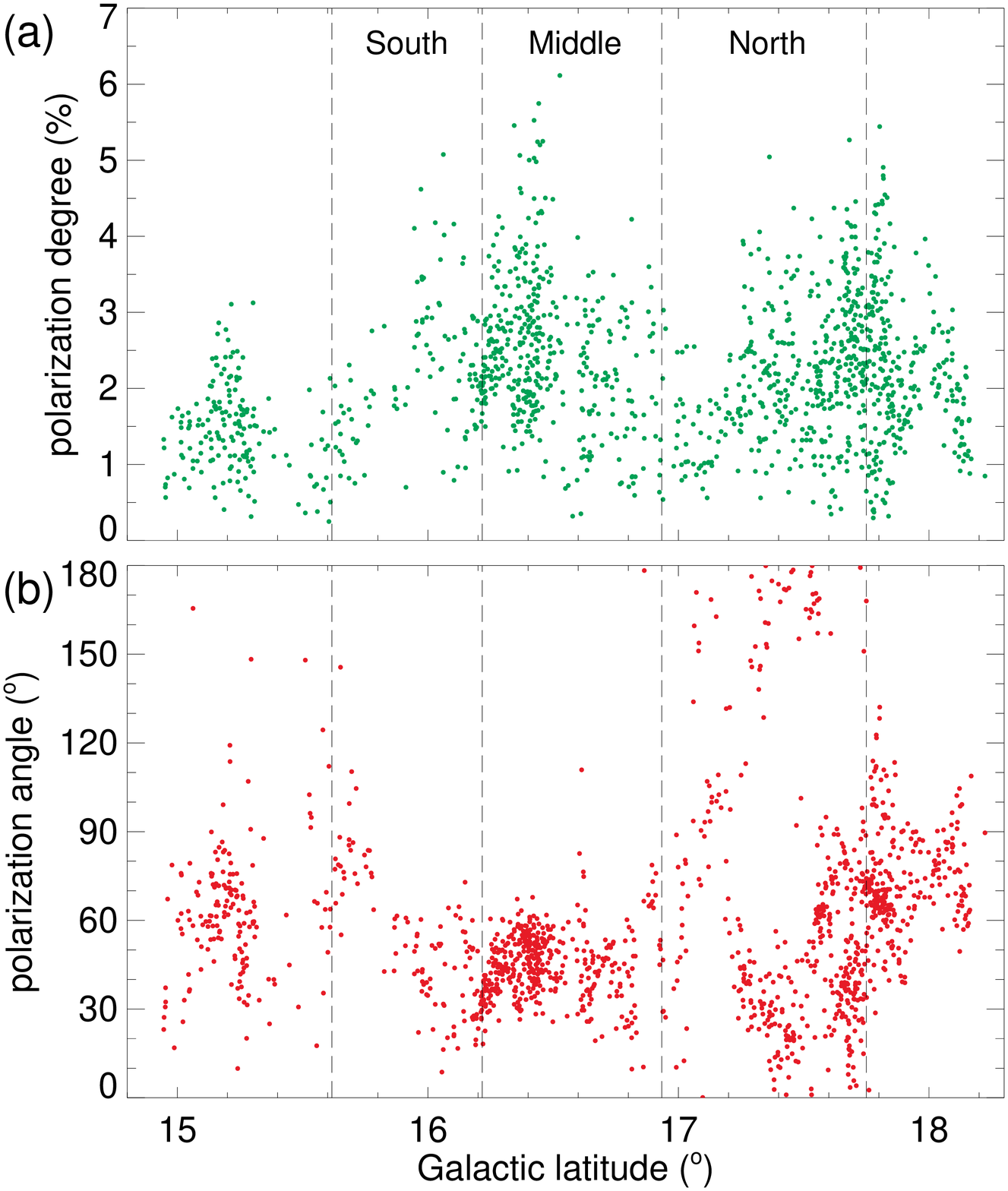}
\caption[]{Distribution of the degree of polarization (upper panel) and polarization angles (lower 
panel) as a function of the Galactic latitude for stars towards the ``main filament''. The vertical
dashed lines delimit the latitudes separating the regions defined in Fig.\,\ref{herschel_gal} for the
main filament.}
\label{angle_x_lat}
\end{figure}

Interesting results can be obtained when we plot the observed polarization data as a
function of the Galactic latitude, as represented in Fig.\,\ref{angle_x_lat}, for the distribution 
of the observed degree of polarization (upper panel) and the polarization angle (lower 
panel). Stars having line-of-sight toward the secondary infrared emission patches 
(boxes ``A'' and ``B'' in Fig.\,\ref{herschel_gal}) were not included in these diagrams.

The vertical dashed-lines delimit the region dominated by each infrared clump. It is
clearly seen that although supposedly belonging to the same interstellar structure, the 
polarimetric characteristics of each clump are distinct. The polarization degree shows a
``wavy'' distribution, ranging from $\sim$0.3\% to little more than 6.0\%. In the region
located southern of the South clump the degree of polarization does not overpass
$\sim$3.0\%, while it rises in the three infrared clumps suggesting a rather good 
correlation between degree of polarization and dust emission.

More interesting is the distribution of polarization angles as a function of the Galactic
latitude. Polarization measured for stars located northern of the North clump present
most of their position angles within the range from $\sim$50\degr\ to 110\degr. Then,
inside the latitudes limiting the vicinity of the North clump we obtained basically any
polarization angles, with a tendency of concentration around $30\degr \pm 20\degr$. 
In the range of latitudes delimiting the Middle clump, we see a narrow distribution centered
around $\sim$40\degr, while in the South clump the position angles show a tendency to 
increase with decreasing Galactic latitude. 

\begin{figure*}
\epsscale{1.1}
\plotone{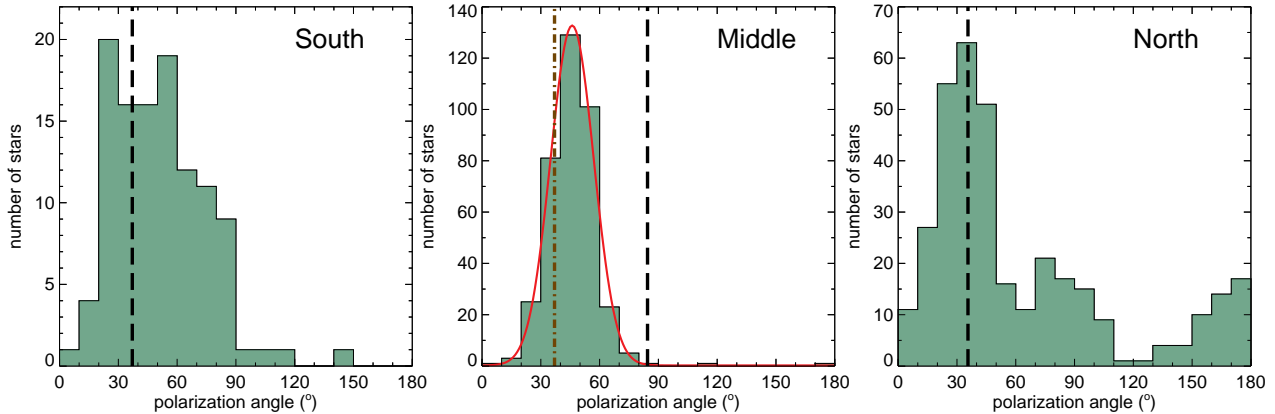}
\caption[]{Distribution of the polarization angles. {\it Left-hand panel} displays the obtained distribution
for stars inside the southern box as defined in Fig.\,\ref{herschel_gal}. The vertical dashed-line indicates the
estimated orthogonal orientation to the main axis of the clump. Middle panel, same as left-hand panel, but
for the middle box. The fitted Gaussian centered at $\sim$46\degr\ and a standard deviation of 11\degr\ shows 
that the observed distribution resembles a normal one, and presents a small dispersion. The vertical
dot-dashed line indicates the orthogonal orientation to the main \lup1\ filament. The {\it right-hand panel}  
displays the observed distribution for stars inside the northern box. The obtained distribution covers almost 
all interval of angles, nevertheless, the main peak in the distribution coincides with the direction perpendicular 
to the longest axis of the clump.
}
\label{hist_main}
\end{figure*}

Some of these properties are better seen in the histograms introduced in 
Fig.\,\ref{hist_main}. The distributions were obtained using stars having lines-of-sight 
through the regions delimited by the rectangles shown in Fig.\,\ref{herschel_gal} only. The 
vertical dashed-lines indicate the visual estimated position angle of the normals to the infrared 
clumps, which for both, the North and South clumps, are basically the 
same as to the whole \lup1 filament, while for the Middle clump, it is tilted by almost 50\degr\
with relation to the former one. Right-hand panel, gives the distribution 
obtained for stars in the North clump region, and as anticipated by the observed distribution 
of polarization angles as function of
the Galactic latitude, the full range of angles are somehow encompassed, 
nevertheless with a clear component almost centered perpendicularly to the infrared 
clump long axis. Middle panel, displays the obtained distribution for stars with lines-of-sight
through the Middle clump region, and  the distribution resembles a normal distribution as 
evidenced by the fitted Gaussian (center at $\sim$46\degr\ and a quite narrow width 
$\sigma = 11\degr$). The obtained angle for the center of this distribution is basically 
identical to the one obtained for the full sample, that is, almost
perpendicular to the long axis of \lup1, however shifted about 40\degr\ from the 
supposed normal to the local infrared clump. Finally, left-hand panel, displays 
the obtained distribution for stars with line-of-sight through the South clump region. Although 
the observed polarization angles do not cover the full range like what was obtained for the 
North clump, this region presents a broad distribution probably due to a merger of different 
magnetic field components.

\subsection{The optical polarization on the infrared diffuse patches}

\begin{figure}
\epsscale{1.1}
\plotone{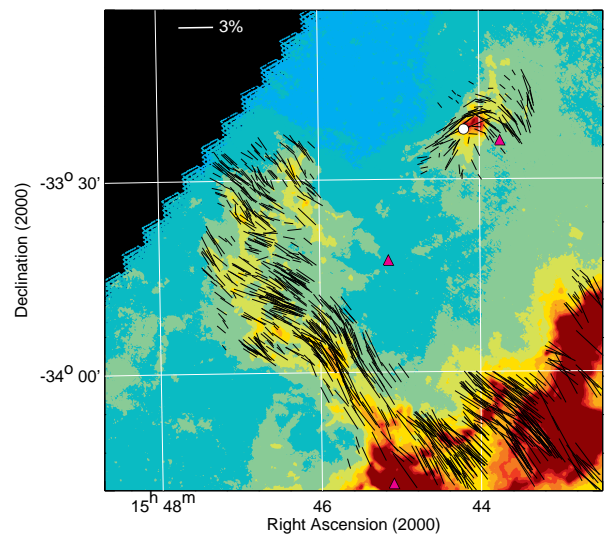}
\caption[]{Detail of the distribution of the linear polarization vectors overlaid on the 
{\it Herschel}/SPIRE 350\,$\mu$m emission map. The symbols have the same meaning
as in Fig.\,\ref{iras_map}. The black area at the upper left corresponds 
to a region not covered by the {\it Herschel} survey.}
\label{herschel_detail_1}
\end{figure}

As mentioned before, our polarimetric survey also covered two adjoining low infrared emission 
clouds at the \lup1 region. The one delimited in Fig.\,\ref{herschel_gal} by the rectangle 
designed by ``A'' corresponds to the northern part of the secondary filament also labeled 
``A'' by \citet[][see their Fig.\,3]{MAA14}. As noted by these authors, this secondary filament is 
perpendicularly aligned to the main \lup1 filament. Fig.\,\ref{herschel_detail_1} gives a detailed 
map of the observed polarization overlaid on the {\it Herschel}/SPIRE 350\,$\mu$m dust emission
for this filament and for the other small diffuse infrared clump delimited by the rectangle designed 
``B'' in Fig.\,\ref{herschel_gal}. The similarity between the features shown by the distribution 
of dust emission and the orientation of the polarization vectors is remarkable, the former 
seems to follow perfectly the field configuration.

\citet{MAA14} have already noted that the extension of the secondary filament A south of the
main \lup1 dust cloud (not covered by our data), was aligned to their submm polarization 
vectors. Our results reinforce their suggestion that \lup1 consists of a dominant filament
running orthogonally to the mean magnetic field, surrounded by secondary filaments that run
parallel to their local field directions. This is in agreement with the recent results obtained by
\citet{AAA14} that diffuse interstellar structures are preferentially aligned with the magnetic field,
while dense molecular clouds appear orthogonally oriented to the magnetic fields.

A question that comes to mind is if secondary filament A is nourishing the main dust filament. 
Unfortunately, our data does not allow to go further to this point, however, unless the
material composing these filaments are distributed on the plane-of-sky, a detailed kinematical 
investigation of them would be interesting. 

While no prestellar object has been so far found in the surveyed area covering the diffuse
filament A, an unbound core and a Class I object were found associated with patch B, the former 
sits in the densest part of the infrared emission observed for this patch. The interesting feature 
observed, for the magnetic field in the region, is its elbow like structure. The
reason for this abrupt bending in the local magnetic field may be associated to the physical
processes leading to the formation of the two young stellar objects found in this area. Further 
higher resolution observations are needed in order to verify that.

\subsection{Angular Dispersion Function, Magnetic Field Strength, and cloud stability}

\subsubsection{The angular dispersion function}

Our polarization map provides the quality and angular resolution to apply the method
of \citet{HKD09} for estimating the plane-of-sky component of the magnetic field acting on
the \lup1 molecular cloud. In their method they use the difference in angle, 
$\Delta\Phi({\ell}) \equiv \Phi(\mathbf{x}) - \Phi(\mathbf{x + \ell})$, between the $N(\ell)$ pairs of 
vectors separated by displacements $\ell$, to compute the angular dispersion function, ADF, 
defined as:

\begin{equation}
\langle \Delta \Phi^2 (\ell)\rangle^{1/2} \equiv \sqrt{\frac{1}{N(\ell)} \sum_{i=1}^{N(\ell)} [ \Phi(\mathbf{x}) 
- \Phi(\mathbf{x + \ell})]^2}.
\label{adf_eq}
\end{equation}

They consider that the magnetic field is composed of a large-scale structured field, 
$\mathbf{B}_0(\mathbf{x})$, and a turbulent component, $\mathbf{B}_t(\mathbf{x})$, and because 
the former is a smoothly varying quantity, its contribution to the dispersion function should increase 
linearly with $\ell$ for small distances. That is, they show that the square of the ADF can be 
approximated by

\begin{figure*}
\epsscale{1.1}
\plotone{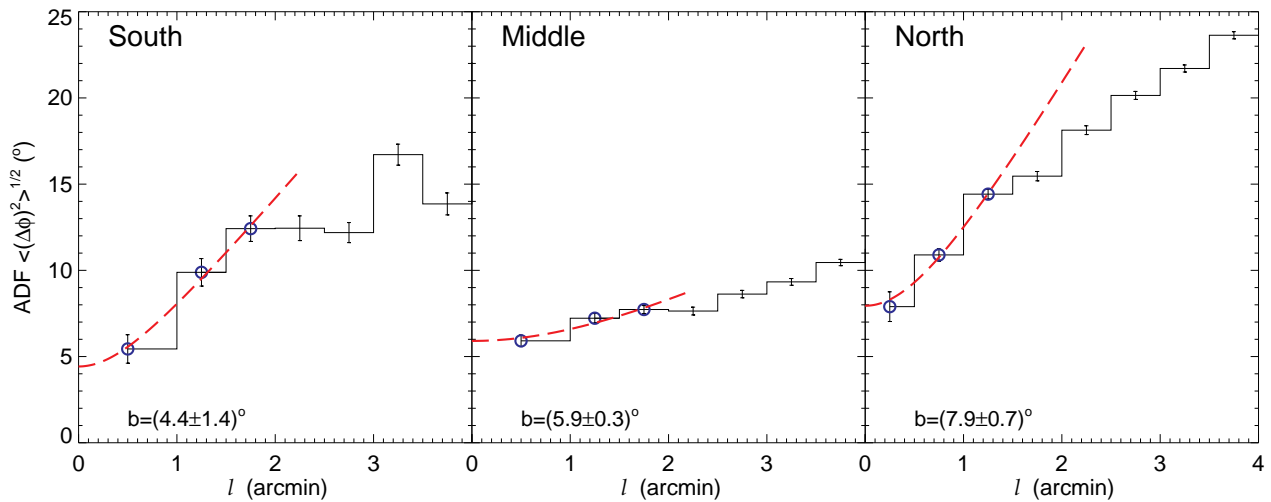}
\caption[]{Plot of the angular dispersion function, $\langle \Delta \Phi^2 (\ell)\rangle^{1/2}$, for the
three clumps of the main \lup1 molecular cloud filament. We considered only polarization vectors 
inside the rectangles shown in Fig.\,\ref{herschel_gal} in calculating these functions. In order to
reduce the errors, indicated by the vertical bars, the first bin in the diagrams for the South and 
Middle clumps were adopted as having twice the size ($1\arcmin$) of the others bins. The best-fit to 
the three first data points are shown by the dashed curves. The turbulent contribution to the total 
angular dispersion is determined by the zero intercept of the best-fit  at $\ell = 0$, and the obtained
value is given at the lower left-hand corner of each panel. The measurement 
uncertainties, that is, the term $\sigma_M (\ell)$ in the right-hand side of 
Eq.\,\ref{adf_fit}, were removed prior to operating the fits to the corresponding data sets.}
\label{adf}
\end{figure*}

\begin{equation}
\langle \Delta \Phi^2 (\ell)\rangle_{\rm tot} \simeq b^2 + m^2\ell^2 + \sigma^2_M(\ell)
\label{adf_fit}
\end{equation}
where $ \sigma_M(\ell)$ represents the contribution due to measurements uncertainties on the
polarization angles and $b$ the turbulent contribution to the angular dispersion, expected to 
maintain constant, as long as $\ell$ is larger than the correlation length characterizing  
$\mathbf{B}_t(\mathbf{x})$.

Figure\,\ref{adf} displays the obtained ADF for the three clumps of the main \lup1 filament. In 
computing that we used only polarization data for stars inside the boxes shown in 
Fig.\,\ref{herschel_gal}. At first glance we note that the turbulent contribution, $b$, in all
three clumps is rather small implying that the ratio of the turbulent to large-scale magnetic
field strength, given by \citep[see,][]{HKD09} 

\begin{equation}
\frac{\langle B_t^2\rangle^{1/2}}{B_0} = \frac{b}{\sqrt{2 - b^2}}
\label{bmag_ratio}
\end{equation}
is significantly less than 1 (see column 3 of Table\,\ref{tab_results}), indicating that turbulence 
is less important than the magnetic fields in these parts of the cloud.

The strength of the plane-of-sky component of the magnetic field may be estimated by
a modified version of the classical method proposed by \citet[][hereafter CF]{CF53}. These
authors assumed equipartition between kinetic and perturbed magnetic energies to assert that 
the dispersion in polarization angles, $\delta\theta$, together with the dispersion in velocity 
along the line-of-sight, $\delta V$, can reveal information about the magnitude of the magnetic 
field. Following the original derivation of CF, an equation for the mean value of the plane-of-sky 
component of the magnetic field can be written as 

\begin{equation}
B_{\rm pos}  =  Q \sqrt{4\pi\rho} \,\frac{\delta V}{\delta \theta}
\label{bpos_def}
\end{equation}
where $\rho$ is the mass density, and $Q$ is a scale factor that equals to unity in the CF model,
however effects not included in the original formula, such as inhomogeneity and averaging several
turbulent cells, among others \citep[see for instance,][]{Zw96}, tend to overestimate the value of the 
field strength. Studies based on simulations suggest that $Q \approx 0.5$ can yield reliable values
of $B_{\rm pos}$, provided the dispersion in angle is less than $\sim$25\degr 
\citep[e.g.,][]{OSG01,HZM01,PGD01}. Adopting $Q = 0.5$ and 
using $\rho = mn_{H_2}$ and $\Delta V = \sqrt{8 \ln 2}\, \delta V$, where $n_{H_2}$ is the molecular 
hydrogen density in molecules cm$^{-3}$ and $\Delta V$ the FWHM line width in km\,s$^{-1}$, we 
obtain \citep{CNW04}

\begin{equation}
B_{\rm pos}  \approx  9.3 \sqrt{n_{H_2}} \,\,\frac{\Delta V}{\delta \theta}
\label{bpos_inter}
\end{equation}
and using the identity $b^2 \approx 2\, \delta\theta^2$, that follows from 
Eq.\,(\ref{bmag_ratio}) when $b^2 \ll 2$  \citep{HKD09},

\begin{equation}
B_{\rm pos} = 9.3 \left ( \frac{2\,n_{H_2}}{\rm cm^{-3}} \right )^{1/2} \left ( \frac{\Delta V}{\rm km\,s^{-1}} \right )
\left ( \frac{b}{1\degr} \right )^{-1} \,\, \mu{\rm G}
\label{bpos_final}
\end{equation}

\subsubsection{Gas density and velocity line width}

In order to obtain the plane-of-sky component of the large-scale magnetic field we need to know 
the values of the gas density and velocity line width toward the investigated line-of-sight. To estimate 
the gas density we assume some hypothesis that will be responsible for the main uncertainty in the 
obtained value of the strength of the plane-of-sky component of the magnetic field. Visual extinctions, 
$A_V$, may be converted to molecular hydrogen column density by using the cannonical relation 
$N(H_2) = 9.4 \times 10^{20} A_V$\,cm$^{-2}$\,mag$^{-1}$ suggested by \citet{BSD78}, and to molecular 
density by assuming a line-of-sight cloud's thickness. However, there is not a precise way to infer the
thickness. Recent investigations conducted in different star forming clouds suggest that filaments are
typically very long, with lengths of $\sim$1\,pc or more, and a possible similar characteristic
width of $\sim$0.1\,pc \citep{AAD11,ADW14}, that is, filamentary clouds may be modelled by an
idealized cylindrical filament with radial density and column density profiles.

The mean visual extinction towards the observed stars in each of the three clumps' areas was 
estimated using the extinction map based on the 2MASS PSC created by 
\citet{DK11}\footnote{http://darkclouds.u-gakugei.ac.jp} and is given in column 4 of Table\,\ref{tab_results}. It 
must be noted that many lines-of-sight returned negative values for the visual extinction, and 
that these data values were not considered in our averaging process. Also, many observed stars 
present degree of polarization much higher than the one expected by the optimum
alignment relation $P_V \approx 3.0 \,A_V$\,\%\,mag$^{-1}$, suggesting that the visual extinction 
obtained from \citeauthor{DK11}'s map, for these lines-of-sight, may be slightly underestimated.
Another sign that the visual extinction retrieved from \citeauthor{DK11}'s map for our stellar 
sample may be underestimated for some lines-of-sight is that \citet{FR02}, in an investigation of 
the surroundings of \lup1, found that stars towards directions having {\it IRAS} 100\,$\mu$m 
emission above 28\,MJy\,sr$^{-1}$, which is the case for the majority of the stars observed in 
this work, present a minimum color excess of $E(b - y) \approx 0.1$\,mag, corresponding to 
$A_V \approx 0.4$\,mag.

To estimate the molecular density we adopted a distance of 150\,pc to the clumps and the width of 
the boxes shown in Fig.\,\ref{herschel_gal} as the thickness of the absorbing material. The obtained 
densities are given in column 5 of Table\,\ref{tab_results}. Although the uncertainties in the 
estimated column density and adopted cloud thickness, the values we obtained for the molecular 
density seem reasonable when compared with the ones obtained in the Taurus
molecular clouds. If we accept that the same physical conditions prevailing 
in those clouds are also valid in \lup1, then we would expect a minimum 
$n(H_2) \approx 375$\,cm$^{-3}$ for lines-of-sight where $^{12}$CO is detected 
\citep[see,][]{PGC10}, meaning that in the worst case, the obtained values give the lower limit for 
the molecular hydrogen density in the surveyed areas.

Unlike the molecular density, the velocity line width is a quantity that does not depend on a priori 
assumptions, however, the choice of which data to use contribute to uncertainties in the final
determination of the strength of the plane-of-sky component of the magnetic field. The Lupus
molecular clouds were mapped a number of times in different molecular lines, however, in general,
none of the data provided in the literature covers exactly our studied regions. For example, 
\citet{TDM96}  obtained $\Delta V = 1.9$\,km\,s$^{-1}$ for $^{13}$CO $J =1 \rightarrow 0$ at the 
peak of integrated intensity for \lup1, and \citet{MCM86} give $\Delta V = 3.55$\,km\,s$^{-1}$ at 
($l =338\fdg5$, $b = 17\fdg5$) and $\Delta V = 2.66$\,km\,s$^{-1}$ at ($l = 339\fdg0$, $b =16\fdg0$), 
two direction towards our mapped area, for $^{12}$CO $J=1 \rightarrow 0$. In this work we 
opted by using the $^{13}$CO $J = 2 \rightarrow 1$ line widths obtained by \citet{TLP09}. They 
observed three lines-of-sight towards the northern clump, eight towards the middle clump, and 
three towards the southern clump, respectively. It must be noted that these measurements were 
gathered for high column density lines-of-sight in these clumps, while we are surveying the outskirts
of the clumps. For each region we adopted the mean value of the velocity line width as given in 
column 6 of Table\,\ref{tab_results}. \citet{TLP09} also provide $^{12}$CO $J = 4 \rightarrow 3$
line width for the same lines-of-sight. The mean values of these lines differ from the ones adopted
by 10-15\%, being larger for the northern clump and smaller for the others two.

The obtained plane-of-sky magnitudes of the large-scale magnetic field are given in column 
7 of Table\,\ref{tab_results}. We note, however, that these values should be taken as an estimation 
only of the actual magnetic field strength, due to uncertainties in the adopted quantities. Although 
the error is dominated by uncertainties in the determination of the molecular hydrogen density 
$n_{H_2}$, we must take into account also the uncertainties in measured line width, the 
estimated $b$ parameter, and the scale factor $Q$, assumed as equal to 0.5.  

\begin{deluxetable*}{lccccccc}
\centering
\tablecaption{Obtained physical parameters for the three clumps in the main \lup1 filament (see text
for details)}
\tablenum{2}
\label{tab_results}
\tablewidth{350pt}
\tablehead{\colhead{Clump} & \colhead{$b$} & \colhead{$\langle B_t^2\rangle^{1/2}/B_0$} & 
\colhead{$\langle A_V \rangle ^{\rm a}$} & \colhead{$n_{H_2}$} & \colhead{$\Delta V^{\rm b}$} &
\colhead{$B_{\rm pos}$} & \colhead{$\lambda$} \\ 
\colhead{} & \colhead{(\degr)} & \colhead{} & \colhead{mag} & \colhead{(cm$^{-3})$} & 
\colhead{(km\,s$^{-1}$)} & \colhead{($\mu$G)} & \colhead{ } } 
\startdata
North & $7.9\pm0.7$ & $0.10\pm0.01$ & 1.80 & 418 & 2.2 & \phn75 & 0.057 \\
Middle & $5.9\pm0.3$ & $0.08\pm0.01$ & 1.93 & 466 & 2.3 & 111 & 0.041 \\
South & $4.4\pm1.4$ & $0.05\pm0.02$ & 2.13 & 540 & 2.7 & 188 &  \phn0.027
\enddata
\tablenotetext{a}{\,\,Based on the extinction map created by \citet{DK11}}
\tablenotetext{b}{\,\,$^{13}$CO $J = 2 \rightarrow 1$ line width from \citet{TLP09}}
\end{deluxetable*}

\subsubsection{Cloud stability}

It is also interesting to investigate the stability of the studied clumps. According to \citet{NN78} 
the critical value for the mass that can be supported by a magnetic flux $\Phi$ is $M_{B,{\rm crit}}
= \Phi/2\pi G^{1/2}$. The ratio of the mass to the magnetic flux is then a crucial parameter that
can supply information on the support and stability of the interstellar clouds. A dimensionless
parameter $\lambda \equiv (M/\Phi)_{\rm actual}/(M/\Phi)_{\rm crit}$, can be estimated using the 
molecular hydrogen density and the magnetic field strength by \citep{CNW04}

\begin{equation}
\lambda = \frac{(M/\Phi)_{\rm observed}}{(M/\Phi)_{\rm crit}} = \frac{mNA/BA}{1/2\pi\sqrt{G}} =
7.6 \times 10^{-21} \frac{N_{\parallel}(H_2)}{B_{\rm tot}}
\label{lambda}
\end{equation}
where $N_{\parallel}$ is the column density along the magnetic flux tube in units of cm$^{-2}$ and  
$B_{\rm tot}$ is the total magnetic field strength given in $\mu$G. As demonstrated by \citet{HC05},
$\langle N_{\rm obs}/B_{\rm pos} \rangle = 3 \langle N_{\parallel}/B_{\rm tot}\rangle$, i.e., using 
statistical arguments they show that in average the observed ratio between column density and
the plane-of-sky magnetic field overestimates the ratio used in Eq. (\ref{lambda}) by a factor of three.
Thus, we can use our previous estimates of molecular column density and plane-of-sky magnetic field
to estimate the $\lambda$ parameter, provided we divide it by 3.

The obtained results for all three clumps in the main filament of \lup1 indicate that their inferred
mass-to-magnetic flux ratios are subcritical (see last column of Table\,\ref{tab_results}). That means, 
in large scale, this molecular cloud is magnetically supported as would be expected for a filament 
perpendicularly aligned with the field lines. Nevertheless, the mass of the main filament of \lup1 may 
be increasing due to a possible longitudinal mass movement of the material in the secondary infrared 
interstellar patches that seems to be channeling into the main filament along the field lines.

In spite of the fact that the parameter $b$, and consequently the ratio of the turbulent to the 
large-scale magnetic field strength, and the mass-to-magnetic flux ratio obtained for all 
three clumps being rather similar, the large-scale magnetic field morphology
in these clumps are not. As already noticed, these differences are clearly observed in the histograms
given in Fig.\,\ref{hist_main}, and are also clear in the ADF plots shown in Fig.\,\ref{adf}.
The large-scale structure is described in Equation\,(\ref{adf_fit}) by the term $m\ell$, which is much
steeper for the North clump and less for the Middle clump. 

It is also interesting to note that a large fraction of the cores and young stellar object found in \lup1 are 
located in these clumps, and that their distribution is not uniform (see Fig.\,\ref{iras_map}). Basically all 
objects found in the North clump seem to be unbound objects, that is, objects having ratio of the core mass 
to its critical Bonnor-Ebert mass less than 1, suggesting that they may or may not form stars. On the other 
hand, in the Middle clump, apart from some supposedly unbound cores, there are also several prestellar 
(bounded) cores and a couple of more evolved objects (Class 0 and I). The South clump seems to be the 
most evolved one, containing several Class I and II objects. All together, it suggests a sequence where the 
southern end of the cloud is more evolved than the middle which in turn is more evolved than the northern 
end. Although we may suspect that there exists a correlation between this fact and the observed 
characteristics of the large-scale magnetic field exerting on the clumps, it is not possible to assert that 
based only in our results.

\section{Conclusions}

We have analyzed the $R$-band imaging linear polarimetry data obtained for background stars
in the region of the \lup1 molecular cloud. Despite the global results are similar to ones obtained in
previous investigations, the higher angular resolution of our data allowed us to obtain new information
not investigated before. Our main results are as follow.

\begin{enumerate}
\item The large scale map of the plane-of-sky magnetic field shows that the main filament of \lup1 is 
normally oriented to the field, reinforcing previous investigations that found the same
\citep[e.g.][]{MG91, RMA98, MAA14}. On the other hand, this map also shows that two small diffuse 
infrared clouds in the neighborhood of the main filament are parallel to the field lines. 
\item When the polarization data are analyzed as a function of the stellar position on the plane-of-sky, 
some interesting features are observed, mainly concerning the distribution of polarization angles as a 
function of the Galactic latitude. The observed characteristic feature seems to be somehow related to 
the clumps of interstellar material. 
\item The ratio of the turbulent to the large-scale magnetic field strength obtained for three major
clumps in \lup1 is significantly less than 1, implying that in this cloud the turbulence is less important 
then the large-scale magnetic fields.
\item The estimated strength of the plane-of-sky component of the large-scale magnetic field ranges 
from $\sim$70 to $\sim$200\,$\mu$G for the North to the South clumps, respectively.
Nevertheless, the reader should bear in mind that these values are very sensitive to the adopted 
quantities characterizing the properties of the probed interstellar medium.
\item There is evidence that the clumps are magnetically supported on large scales. This is in 
agreement with the observational fact that the gas and dust composing the main structure of
\lup1 have collapsed along the field lines to form a filament. There is also evidence that this process
is still ongoing by the accretion of material along field lines like what is suggested by the result
obtained for one of the observed diffuse infrared patches. 
\end{enumerate}

The results introduced here and the fact that \lup1 is currently undergoing a large star formation 
event \citep{RBS13}, make this cloud an interesting site for future investigations on the
importance of magnetic fields in the star formation process.

\acknowledgements
We thank the referee for her/his insightful and valuable comments that substantially 
improved our paper.
We thank the staff of the Observat\'orio do Pico dos Dias 
(LNA/MCTI, Brazil) for their hospitality and invaluable help during our observing 
runs. 
This research has made use of the NASA/IPAC Infrared 
Science Archive, which is operated by the Jet Propulsion Laboratory, 
California Institute of Technology, under contract with the National 
Aeronautics and Space Administration through the use of data products 
from the Two Micron All Sky Survey, which is a joint project of the University of 
Massachusetts and the Infrared Processing and Analysis Center/California Institute 
of Technology, funded by the National Aeronautics and Space Administration and the 
National Science Foundation and from the {\it Infrared Astronomical Satellite} which was 
a joint project of the US, UK and the Netherlands.
This research has made use of data from the Herschel Gould Belt survey (HGBS) project. The 
HGBS is a Herschel Key Programme jointly carried out by SPIRE Specialist Astronomy Group 3 
(SAG 3), scientists of several institutes in the PACS Consortium (CEA Saclay, INAF-IFSI Rome 
and INAF-Arcetri, KU Leuven, MPIA Heidelberg), and scientists of the Herschel Science 
Center (HSC).
This research has also made use of the Digitized Sky Survey produced
at the Space Telescope Science Institute under U.S. Government grant 
NAG W-2166, which is based on photographic data obtained using the UK Schmidt 
Telescope and the Palomar Sky Survey. 
We have also made extensive use of NASA's Astrophysics Data System (NASA/ADS) 
and the SIMBAD database, operated at CDS, Strasbourg, France. 
We are grateful to Drs. A. M. Magalh\~aes and A. Pereyra for supplying us 
with the polarimetric unit and the software used for data reductions. 
This research has been partially supported by CNPq and FAPEMIG.

{\it Facilities:} \facility{LNA: 1.6\,m and BC0.6\,m}


\begin{thebibliography}{56}
\expandafter\ifx\csname natexlab\endcsname\relax\def\natexlab#1{#1}\fi

\bibitem[{{Alves} \& {Franco}(2006)}]{AF06}
{Alves}, F.~O. \& {Franco}, G.~A.~P. 2006, \mnras, 366, 238

\bibitem[{{Alves} {et~al.}(2008){Alves}, {Franco}, \& {Girart}}]{AFG08}
{Alves}, F.~O., {Franco}, G.~A.~P., \& {Girart}, J.~M. 2008, \aap, 486, L13

\bibitem[{{Alves} {et~al.}(2014){Alves}, {Frau}, {Girart}, {Franco}, {Santos},
  \& {Wiesemeyer}}]{AFG14}
{Alves}, F.~O., {Frau}, P., {Girart}, J.~M., {et~al.} 2014, \aap, 569, L1

\bibitem[{{Andr{\'e}} {et~al.}(2014){Andr{\'e}}, {Di Francesco},
  {Ward-Thompson}, {Inutsuka}, {Pudritz}, \& {Pineda}}]{ADW14}
{Andr{\'e}}, P., {Di Francesco}, J., {Ward-Thompson}, D., {et~al.} 2014,
  Protostars and Planets VI, 27

\bibitem[{{Andr{\'e}} \& {Saraceno}(2005)}]{AS05}
{Andr{\'e}}, P. \& {Saraceno}, P. 2005, in ESA Special Publication, Vol. 577,
  ESA Special Publication, ed. A.~{Wilson}, 179--184

\bibitem[{{Arzoumanian} {et~al.}(2011){Arzoumanian}, {Andr{\'e}}, {Didelon},
  {K{\"o}nyves}, {Schneider}, {Men'shchikov}, {Sousbie}, {Zavagno}, {Bontemps},
  {di Francesco}, {Griffin}, {Hennemann}, {Hill}, {Kirk}, {Martin}, {Minier},
  {Molinari}, {Motte}, {Peretto}, {Pezzuto}, {Spinoglio}, {Ward-Thompson},
  {White}, \& {Wilson}}]{AAD11}
{Arzoumanian}, D., {Andr{\'e}}, P., {Didelon}, P., {et~al.} 2011, \aap, 529, L6

\bibitem[{{Benedettini} {et~al.}(2012){Benedettini}, {Pezzuto}, {Burton},
  {Viti}, {Molinari}, {Caselli}, \& {Testi}}]{BPB12}
{Benedettini}, M., {Pezzuto}, S., {Burton}, M.~G., {et~al.} 2012, \mnras, 419,
  238

\bibitem[{{Bohlin} {et~al.}(1978){Bohlin}, {Savage}, \& {Drake}}]{BSD78}
{Bohlin}, R.~C., {Savage}, B.~D., \& {Drake}, J.~F. 1978, \apj, 224, 132

\bibitem[{{Cambr{\'e}sy}(1999)}]{Cam99b}
{Cambr{\'e}sy}, L. 1999, \aap, 345, 965

\bibitem[{{Chandrasekhar} \& {Fermi}(1953)}]{CF53}
{Chandrasekhar}, S. \& {Fermi}, E. 1953, \apj, 118, 113

\bibitem[{{Comer{\'o}n}(2008)}]{FC08}
{Comer{\'o}n}, F. 2008, {Handbook of Star Forming Regions, Volume II: The
  Southern Sky, Edited by Reipurth, B.} (ASP Monograph), 295

\bibitem[{{Crawford}(2000)}]{CIA00}
{Crawford}, I.~A. 2000, \mnras, 317, 996

\bibitem[{{Crutcher}(2012)}]{Cr12}
{Crutcher}, R.~M. 2012, \araa, 50, 29

\bibitem[{{Crutcher} {et~al.}(2004){Crutcher}, {Nutter}, {Ward-Thompson}, \&
  {Kirk}}]{CNW04}
{Crutcher}, R.~M., {Nutter}, D.~J., {Ward-Thompson}, D., \& {Kirk}, J.~M. 2004,
  \apj, 600, 279

\bibitem[{{Davis} \& {Greenstein}(1951)}]{DG51}
{Davis}, L.~J. \& {Greenstein}, J.~L. 1951, \apj, 114, 206

\bibitem[{{Dobashi}(2011)}]{DK11}
{Dobashi}, K. 2011, \pasj, 63, 1

\bibitem[{{Dolginov} \& {Mytrophanov}(1976)}]{DM76}
{Dolginov}, A.~Z. \& {Mytrophanov}, I.~G. 1976, \apss, 43, 257

\bibitem[{{Franco}(2002)}]{FR02}
{Franco}, G.~A.~P. 2002, \mnras, 331, 474

\bibitem[{{Franco} {et~al.}(2010){Franco}, {Alves}, \& {Girart}}]{FAG10}
{Franco}, G.~A.~P., {Alves}, F.~O., \& {Girart}, J.~M. 2010, \apj, 723, 146

\bibitem[{{Heiles} \& {Crutcher}(2005)}]{HC05}
{Heiles}, C. \& {Crutcher}, R. 2005, in Lecture Notes in Physics, Berlin
  Springer Verlag, Vol. 664, Cosmic Magnetic Fields, ed. R.~{Wielebinski} \&
  R.~{Beck}, 137

\bibitem[{{Heitsch} {et~al.}(2001){Heitsch}, {Zweibel}, {Mac Low}, {Li}, \&
  {Norman}}]{HZM01}
{Heitsch}, F., {Zweibel}, E.~G., {Mac Low}, M.-M., {Li}, P., \& {Norman}, M.~L.
  2001, \apj, 561, 800

\bibitem[{{Hennebelle} \& {Falgarone}(2012)}]{HF12}
{Hennebelle}, P. \& {Falgarone}, E. 2012, \aapr, 20, 55

\bibitem[{{Heyer} {et~al.}(1987){Heyer}, {Vrba}, {Snell}, {Schloerb}, {Strom},
  {Goldsmith}, \& {Strom}}]{HVS87}
{Heyer}, M.~H., {Vrba}, F.~J., {Snell}, R.~L., {et~al.} 1987, \apj, 321, 855

\bibitem[{{Hildebrand} {et~al.}(2009){Hildebrand}, {Kirby}, {Dotson}, {Houde},
  \& {Vaillancourt}}]{HKD09}
{Hildebrand}, R.~H., {Kirby}, L., {Dotson}, J.~L., {Houde}, M., \&
  {Vaillancourt}, J.~E. 2009, \apj, 696, 567

\bibitem[{{Hoang} \& {Lazarian}(2014)}]{HL14}
{Hoang}, T. \& {Lazarian}, A. 2014, \mnras, 438, 680

\bibitem[{{Hoang} {et~al.}(2015){Hoang}, {Lazarian}, \& {Andersson}}]{HLA15}
{Hoang}, T., {Lazarian}, A., \& {Andersson}, B.-G. 2015, \mnras, 448, 1178

\bibitem[{{Jones} {et~al.}(2015){Jones}, {Bagley}, {Krejny}, {Andersson}, \&
  {Bastien}}]{JBK15}
{Jones}, T.~J., {Bagley}, M., {Krejny}, M., {Andersson}, B.-G., \& {Bastien},
  P. 2015, \aj, 149, 31

\bibitem[{{Lazarian}(2007)}]{L07}
{Lazarian}, A. 2007, \jqsrt, 106, 225

\bibitem[{{Li} {et~al.}(2009){Li}, {Dowell}, {Goodman}, {Hildebrand}, \&
  {Novak}}]{LDG09}
{Li}, H.-b., {Dowell}, C.~D., {Goodman}, A., {Hildebrand}, R., \& {Novak}, G.
  2009, \apj, 704, 891

\bibitem[{{Li} {et~al.}(2013){Li}, {Fang}, {Henning}, \& {Kainulainen}}]{LFH13}
{Li}, H.-b., {Fang}, M., {Henning}, T., \& {Kainulainen}, J. 2013, \mnras, 436,
  3707

\bibitem[{{Lombardi} {et~al.}(2008){Lombardi}, {Lada}, \& {Alves}}]{LLA08}
{Lombardi}, M., {Lada}, C.~J., \& {Alves}, J. 2008, \aap, 480, 785

\bibitem[{{Low} {et~al.}(1984){Low}, {Young}, {Beintema}, {Gautier},
  {Beichman}, {Aumann}, {Gillett}, {Neugebauer}, {Boggess}, \& {Emerson}}]{low}
{Low}, F.~J., {Young}, E., {Beintema}, D.~A., {et~al.} 1984, \apjl, 278, L19

\bibitem[{{Mac Low} \& {Klessen}(2004)}]{MK04}
{Mac Low}, M.-M. \& {Klessen}, R.~S. 2004, Reviews of Modern Physics, 76, 125

\bibitem[{{Magalh\~aes} {et~al.}(1984){Magalh\~aes}, {Benedetti}, \&
  {Roland}}]{MBR84}
{Magalh\~aes}, A.~M., {Benedetti}, E., \& {Roland}, E.~H. 1984, \pasp, 96, 383

\bibitem[{{Magalh\~aes} {et~al.}(1996){Magalh\~aes}, {Rodrigues}, {Margoniner},
  {Pereyra}, \& {Heathcote}}]{AM96}
{Magalh\~aes}, A.~M., {Rodrigues}, C.~V., {Margoniner}, V.~E., {Pereyra}, A.,
  \& {Heathcote}, S. 1996, in ASP Conf. Ser. 97: Polarimetry of the
  Interstellar Medium, 118

\bibitem[{{Matthews} {et~al.}(2014){Matthews}, {Ade}, {Angil{\`e}}, {Benton},
  {Chapin}, {Chapman}, {Devlin}, {Fissel}, {Fukui}, {Gandilo}, {Gundersen},
  {Hargrave}, {Klein}, {Korotkov}, {Moncelsi}, {Mroczkowski}, {Netterfield},
  {Novak}, {Nutter}, {Olmi}, {Pascale}, {Poidevin}, {Savini}, {Scott},
  {Shariff}, {Soler}, {Tachihara}, {Thomas}, {Truch}, {Tucker}, {Tucker}, \&
  {Ward-Thompson}}]{MAA14}
{Matthews}, T.~G., {Ade}, P.~A.~R., {Angil{\`e}}, F.~E., {et~al.} 2014, \apj,
  784, 116

\bibitem[{{McKee} \& {Ostriker}(2007)}]{MO07}
{McKee}, C.~F. \& {Ostriker}, E.~C. 2007, \araa, 45, 565

\bibitem[{{Molinari} {et~al.}(2010){Molinari}, {Swinyard}, {Bally}, {Barlow},
  {Bernard}, {Martin}, {Moore}, {Noriega-Crespo}, {Plume}, {Testi}, {Zavagno},
  {Abergel}, {Ali}, {Anderson}, {Andr{\'e}}, {Baluteau}, {Battersby},
  {Beltr{\'a}n}, {Benedettini}, {Billot}, {Blommaert}, {Bontemps}, {Boulanger},
  {Brand}, {Brunt}, {Burton}, {Calzoletti}, {Carey}, {Caselli}, {Cesaroni},
  {Cernicharo}, {Chakrabarti}, {Chrysostomou}, {Cohen}, {Compiegne}, {de
  Bernardis}, {de Gasperis}, {di Giorgio}, {Elia}, {Faustini}, {Flagey},
  {Fukui}, {Fuller}, {Ganga}, {Garcia-Lario}, {Glenn}, {Goldsmith}, {Griffin},
  {Hoare}, {Huang}, {Ikhenaode}, {Joblin}, {Joncas}, {Juvela}, {Kirk},
  {Lagache}, {Li}, {Lim}, {Lord}, {Marengo}, {Marshall}, {Masi}, {Massi},
  {Matsuura}, {Minier}, {Miville-Desch{\^e}nes}, {Montier}, {Morgan}, {Motte},
  {Mottram}, {M{\"u}ller}, {Natoli}, {Neves}, {Olmi}, {Paladini}, {Paradis},
  {Parsons}, {Peretto}, {Pestalozzi}, {Pezzuto}, {Piacentini}, {Piazzo},
  {Polychroni}, {Pomar{\`e}s}, {Popescu}, {Reach}, {Ristorcelli}, {Robitaille},
  {Robitaille}, {Rod{\'o}n}, {Roy}, {Royer}, {Russeil}, {Saraceno}, {Sauvage},
  {Schilke}, {Schisano}, {Schneider}, {Schuller}, {Schulz}, {Sibthorpe},
  {Smith}, {Smith}, {Spinoglio}, {Stamatellos}, {Strafella}, {Stringfellow},
  {Sturm}, {Taylor}, {Thompson}, {Traficante}, {Tuffs}, {Umana}, {Valenziano},
  {Vavrek}, {Veneziani}, {Viti}, {Waelkens}, {Ward-Thompson}, {White},
  {Wilcock}, {Wyrowski}, {Yorke}, \& {Zhang}}]{MSB10}
{Molinari}, S., {Swinyard}, B., {Bally}, J., {et~al.} 2010, \aap, 518, L100

\bibitem[{{Mouschovias} \& {Ciolek}(1999)}]{MC99}
{Mouschovias}, T.~C. \& {Ciolek}, G.~E. 1999, in NATO ASIC Proc. 540: The
  Origin of Stars and Planetary Systems, ed. C.~J. {Lada} \& N.~D. {Kylafis},
  305

\bibitem[{{Murphy} {et~al.}(1986){Murphy}, {Cohen}, \& {May}}]{MCM86}
{Murphy}, D.~C., {Cohen}, R., \& {May}, J. 1986, \aap, 167, 234

\bibitem[{{Myers} \& {Goodman}(1991)}]{MG91}
{Myers}, P.~C. \& {Goodman}, A.~A. 1991, \apj, 373, 509

\bibitem[{{Nakano} \& {Nakamura}(1978)}]{NN78}
{Nakano}, T. \& {Nakamura}, T. 1978, \pasj, 30, 671

\bibitem[{{Ostriker} {et~al.}(2001){Ostriker}, {Stone}, \& {Gammie}}]{OSG01}
{Ostriker}, E.~C., {Stone}, J.~M., \& {Gammie}, C.~F. 2001, \apj, 546, 980

\bibitem[{{Padoan} {et~al.}(2001){Padoan}, {Goodman}, {Draine}, {Juvela},
  {Nordlund}, \& {R{\"o}gnvaldsson}}]{PGD01}
{Padoan}, P., {Goodman}, A., {Draine}, B.~T., {et~al.} 2001, \apj, 559, 1005

\bibitem[{{Pereyra} \& {Magalh{\~a}es}(2004)}]{PM04}
{Pereyra}, A. \& {Magalh{\~a}es}, A.~M. 2004, \apj, 603, 584

\bibitem[{{Pineda} {et~al.}(2010){Pineda}, {Goldsmith}, {Chapman}, {Snell},
  {Li}, {Cambr{\'e}sy}, \& {Brunt}}]{PGC10}
{Pineda}, J.~L., {Goldsmith}, P.~F., {Chapman}, N., {et~al.} 2010, \apj, 721,
  686

\bibitem[{{Planck Collaboration XXXII}(2014)}]{AAA14}
{Planck Collaboration XXXII}. 2014, ArXiv e-prints:1409.6728

\bibitem[{{Rizzo} {et~al.}(1998){Rizzo}, {Morras}, \& {Arnal}}]{RMA98}
{Rizzo}, J.~R., {Morras}, R., \& {Arnal}, E.~M. 1998, \mnras, 300, 497

\bibitem[{{Rygl} {et~al.}(2013){Rygl}, {Benedettini}, {Schisano}, {Elia},
  {Molinari}, {Pezzuto}, {Andr{\'e}}, {Bernard}, {White}, {Polychroni},
  {Bontemps}, {Cox}, {Di Francesco}, {Facchini}, {Fallscheer}, {di Giorgio},
  {Hennemann}, {Hill}, {K{\"o}nyves}, {Minier}, {Motte}, {Nguyen-Luong},
  {Peretto}, {Pestalozzi}, {Sadavoy}, {Schneider}, {Spinoglio}, {Testi}, \&
  {Ward-Thompson}}]{RBS13}
{Rygl}, K.~L.~J., {Benedettini}, M., {Schisano}, E., {et~al.} 2013, \aap, 549,
  L1

\bibitem[{{Serkowski}(1974)}]{SK74}
{Serkowski}, K. 1974, in {Methods Exper. Phys. Vol.~12A}, ed. N.~{Carleton}
  (Academic, New York), 361

\bibitem[{{Soler} {et~al.}(2013){Soler}, {Hennebelle}, {Martin},
  {Miville-Desch{\^e}nes}, {Netterfield}, \& {Fissel}}]{SHM12}
{Soler}, J.~D., {Hennebelle}, P., {Martin}, P.~G., {et~al.} 2013, \apj, 774,
  128

\bibitem[{{Tachihara} {et~al.}(1996){Tachihara}, {Dobashi}, {Mizuno}, {Ogawa},
  \& {Fukui}}]{TDM96}
{Tachihara}, K., {Dobashi}, K., {Mizuno}, A., {Ogawa}, H., \& {Fukui}, Y. 1996,
  \pasj, 48, 489

\bibitem[{{Tothill} {et~al.}(2009){Tothill}, {L{\"o}hr}, {Parshley}, {Stark},
  {Lane}, {Harnett}, {Wright}, {Walker}, {Bourke}, \& {Myers}}]{TLP09}
{Tothill}, N.~F.~H., {L{\"o}hr}, A., {Parshley}, S.~C., {et~al.} 2009, \apjs,
  185, 98

\bibitem[{{Van Loo} {et~al.}(2014){Van Loo}, {Keto}, \& {Zhang}}]{VKZ14}
{Van Loo}, S., {Keto}, E., \& {Zhang}, Q. 2014, \apj, 789, 37

\bibitem[{{V{\'a}zquez-Semadeni} {et~al.}(2011){V{\'a}zquez-Semadeni},
  {Banerjee}, {G{\'o}mez}, {Hennebelle}, {Duffin}, \& {Klessen}}]{VSB11}
{V{\'a}zquez-Semadeni}, E., {Banerjee}, R., {G{\'o}mez}, G.~C., {et~al.} 2011,
  \mnras, 414, 2511

\bibitem[{{Zweibel}(1996)}]{Zw96}
{Zweibel}, E.~G. 1996, in Astronomical Society of the Pacific Conference
  Series, Vol.~97, Polarimetry of the Interstellar Medium, ed. W.~G. {Roberge}
  \& D.~C.~B. {Whittet}, 486

\end{thebibliography}
\end{document}